\documentclass[aps,prl,nobibnotes,showpacs,twocolumn]{revtex4}

\usepackage{graphicx}% Include figure files
\usepackage{epsfig}

\usepackage{amsmath,amssymb,amsthm}

\usepackage{color,ulem}

\newcommand{\ket}[1]{|#1\rangle}

\newcommand{\bra}[1]{\langle#1|}

\newcommand{\tr}[1]{{\rm tr}\left\{#1\right\}}
\newcommand{\ptr}[2]{{\rm tr}_{#2}\left\{#1\right\}}

\def\T{\mathcal{T}}
\def\wT{{\widetilde{\mathcal{T}}}}
\def\D{\mathcal{D}}
\def\W{\mathcal{W}}
\def\wW{{\widetilde{\mathcal{W}}}}
\def\H{\mathcal{H}}
\def\E{\mathcal{E}}

\def\bea{\begin{eqnarray}}
\def\eea{\end{eqnarray}}

\begin{document}

%%%%%%%%%%%%%%%%%%%%%%%%%%%%%%%%%%%%%%%%%%%%%%%%
\title{Optimal approximate transpose map via quantum designs and its applications to entanglement detection}

\author{ Amir Kalev}
\affiliation{Centre for Quantum Technologies, National University of Singapore, 3 Science Drive 2, 117543, Singapore\\Center for Quantum Information and Control, University of New Mexico, Albuquerque, NM 87131-0001, USA\footnote{Current address}}

\author{Joonwoo Bae}
\affiliation{ Center for Quantum Technologies, National University of Singapore, 3 Science Drive 2, Singapore 117543}

\date{\today}

\begin{abstract}
We show that quantum designs characterize the general structure of the optimal approximation of the transpose map on quantum states. Based on this characterization, we propose an implementation of the approximate transpose map by a measurement-and-preparation scheme. The results show that state-manipulation in quantum two-designs suffices for transpose-based quantum information applications. In particular, we present how these results can be applied to the framework of detecting multipartite entangled states, for instance, when local measurements or interferometry-based experimental approaches are applied.
\end{abstract}


\maketitle

%--------------------------------------------------------------------------------------------------------------------

{\it Introduction}. The transpose map on quantum states has been a unique theoretical tool for the formulation of quantum dynamics. Namely, it characterizes anti-unitary transformations, a class of non-legitimate quantum operations \cite{ref:wigner}, and explains some of impossible tasks for qubit systems such as the universal-NOT operation \cite{ref:werner} or the anti-cloning operation \cite{ref:gisin}. In quantum information theory, the map plays a crucial role in entanglement-based information tasks. In particular, it is closely related to distillable entanglement \cite{ref:nppt} and those quantum states violating Bell's inequalities \cite{ref:peresconjecture}, as well as generally a method of detecting entanglement \cite{ref:peres, ref:mpr}, see also the review Ref. \cite{ref:ent-rev1}. 

While impossible is to realize the transpose map in a laboratory from the principle, still what would be possible to have in experiment is its usefulness in entanglement theory such as detecting entangled states: a quantum operation that approximates the partial transpose can be exploited for entanglement detection \cite{ref:ek}. To realize the practical purpose in experiment i.e. toward detecting entangled states in practice based on the approximations, the followings are needed, both i) theoretic methods of approximating the transpose map such that its usefulness is not lost and ii) their implementation schemes devised in a way that they are feasible with current experimental technologies. Clearly, this is also of fundamental interest in its own right to find a physical approximation to the transpose map and its realization within quantum theory.

In fact, for the transpose map, a particular method, the so-called structural physical approximation\cite{ref:spa}, provides an optimal approximation in the sense that it attains the maximal fidelity that can be achieved within quantum theory \cite{ref:werner, ref:bus}. In the recent years, it was shown that the approximate transpose can generally be constructed in experiment by a measurement-and-preparation scheme, and thus its feasibility with present-day technologies has been shown \cite{ref:korbicz}. Proof-of-principle demonstrations in experiment are then reported with photonics quantum devices \cite{ref:lim}. Despite the rapid developments in realizations and applications to entanglement detection, however, the general structure to characterize the precise relation between the measurements and the approximate transpose is far from being understood. What measurements would optimally approximate the transpose, and how are they related each other? This amounts, on the fundamental side, to find how the transpose map is approximated by legitimate quantum operations, and then on the practical side, to devise and improve further implementation schemes for particular applications, beyond the proof-of-principle demonstrations.

In this work, we show that the optimal approximation of the transpose map can be written in terms of 
quantum two-designs. This mathematical relation has an operational meaning when the designs correspond to measurements, for which the complete set of mutually unbiased bases (MUB) \cite{ref:ivanovic,ref:wootters89} and symmetric, informationally complete probability-operator measurements (SIC POMs) \cite{ref:sic}, are two well known examples \cite{ref:des1}. Based on this mathematical structure, we propose an implementation scheme of the approximate map based on SIC POM, which is a design with a minimal cardinality. The proposed scheme works in a trace-preserving manner, not relying on post-selected data. We then show how this construction can be applied to detecting multipartite entangled states. In particular, the detection scheme provides the possibilities to interferometry-based experimental approaches, that would be natural realizations of detecting entanglement. Our results show that state-manipulation over quantum two-designs generally suffices to realize transpose-based quantum information applications.

{\it The approximate map and quantum designs}. Let us first show the explicit relation between a spherical two-design and the optimal physical approximation of the transpose map. For this purpose, we summarize three known technical results regarding the approximate transpose \cite{ref:spa}, the Choi-Jamio{\l}kowski (CJ) isomorphism \cite{ref:iso}, and quantum designs \cite{ref:des1, ref:des2, ref:des3, ref:des4, ref:sic, ref:scott06}. 

First, let $\T$ denote the transpose map for operators acting on a $d$-dimensional Hilbert space, $\T[\ket{i}\!\bra{j}]=\ket{j}\!\bra{i}$ where the kets and bras are in the computational basis. The approximation of $\T$,
\bea
\wT = \frac{1}{d+1}\T + \frac{d}{d+1}\D,~\mathrm{where} ~ \D[\rho] = \tr{\rho}\frac{\openone}{d},
\label{eq:appt}
\eea
provides the highest average fidelity that can be achieved by quantum operations \cite{ref:spa}.

The approximate map $\wT$ can be characterized by a bipartite quantum state, through the isomorphism \cite{ref:iso} established between quantum operations and bipartite quantum states.  Let $S(\H_d)$ denote the set of quantum states in a $d$-dimensional Hilbert space. The one-to-one correspondence between quantum operations $\E: S(\H_d)\rightarrow S(\H_d) $ and quantum states $\chi \in S (\H_d \otimes \H_d)$ is formulated as follows: For all states $\chi_\E = (\mathcal{I} \otimes \E ) [|\phi^{+}\rangle\!\langle \phi^{+} | ] $ called CJ state where $|\phi^{+}\rangle = \sum_{i=1}^d |ii\rangle /\sqrt{d}$, there exist quantum operations $\E [\rho] = d~\ptr{\chi_\E  ( \T[\rho_{A}]\otimes\openone_B ) }{A}$, and vice versa. The CJ state gives the complete characterization of its corresponding quantum operation. Using these relations, one finds that the CJ state corresponding to $\wT$ is given by  \cite{ref:korbicz}, 
\bea
\rho_{\widetilde{\mathcal T}}=\frac{1}{d(d+1)}\bigl(\openone+V\bigr)=\frac{2}{d(d+1)} P_{\mathrm{sym}},
\label{eq:state-appt-sym}
\eea
where $V$ is the swap operator, $V\ket{\psi}\ket{\phi}=\ket{\phi}\ket{\psi}$, and  $P_{\mathrm{sym}}$ is the projection onto the symmetric subspace of $S(\H_d\otimes \H_d)$. We note that $\rho_\wT$ is separable.  However it is generally not straightforward to find separable decompositions of separable states, and it is even harder to find the decomposition of a minimal cardinality.

To write the separable decomposition of $\rho_{\widetilde{\mathcal T}}$ we use the result form quantum design theory where it is known that the projector $P_{\mathrm{sym}}$ defines a spherical two-design. Any set of states $\{ |x_{k}\rangle \}_{k=1}^{N}$ that fulfills 
\bea
P_{\mathrm{sym}}=\frac{1}{N} \sum_{k=1}^{N} |x_k\rangle \langle x_k |\otimes|x_k\rangle \langle x_k |
\label{eq:spherical}
\eea
in $S(\H_d\otimes\H_d)$, is called a spherical two-design. If moreover, $\sum_{k} |x_k\rangle \langle x_k |{\propto}\openone$, then the design is called a {\it coherent} design.  Having collected all these results, it  follows that the approximate map can be written as
\bea
\widetilde{\mathcal T}[\rho] = \frac{1}{N}\sum_{k=1}^{N} \langle x_k | \rho | x_k \rangle |x_{k}^{*} \rangle \langle x_{k}^{*}|,~~\forall\rho\in S(\H_d), \label{eq:apt}
\eea
where the star denotes a complex conjugation.

Eqs. (\ref{eq:state-appt-sym})-(\ref{eq:apt}) show that both the approximate map $\wT$ and its corresponding CJ state $\rho_\wT$ can be written in terms of quantum spherical two-designs. If a design is a coherent design, then Eq.~(\ref{eq:apt}), has an operational meaning: A measurement and preparation in coherent spherical two-designs optimally approximate the transpose map and thus that state-manipulation in coherent two-designs, when exist, suffices to realize transpose-based quantum information applications.

As mentioned before, two well-known instances of coherent spherical two-designs are the collection of $(d+1)$ MUB, denoted by $\{|b_{j}\rangle \}_{j=1}^{d(d+1)}$,  and SIC states, $\{|s_{j}\rangle \}_{j=1}^{d^2}$, which are used to construct SIC POMs \cite{ref:mubsic}. Measurements in MUB and SIC POMs are in fact those simplest settings often applied in experiment, e.g. for quantum state tomography. Then, from the general construction of Eq.~(\ref{eq:apt}), it follows that
\bea
\widetilde{\mathcal T} [\rho] &=& \sum_{j=1}^{d(d+1)} \tr{\frac{| b_j \rangle\langle b_j|}{d(d+1)}\rho} | b_{j}^{*} \rangle \langle b_{ j }^{*} | \label{eq:mubt} \\
&=&  \sum_{j=1}^{d^2} \tr{ |s_j \rangle\frac{1}{d} \langle s_j | \rho } |s_{j}^{*} \rangle \langle s_{j}^{*} |,\label{eq:sict}
\eea
i.e. the approximate map could be realized by a measurement-and-preparation scheme using either MUB, Eq.~(\ref{eq:mubt}), or  SIC states, Eq.~(\ref{eq:sict}).

As a simple illustration of the above result, let us consider the implementation of $\wT$ on a qubit state $\rho$ through a measurement-and-preparation scheme in the three MUB of a qubit. To implement $\wT [\rho]$ the system is measured randomly in one of the eigenbases of  $\sigma_x$, $\sigma_y$ or $\sigma_z$. If detection happens in the $\sigma_y$ basis, and the state $|b_y\rangle$ for $b=0,1$, is detected, then one prepares the state $|\bar{b}_{y}\rangle$ with bar denotes  the logical NOT operation. For measurements in the other two bases the states on which detection events have happened are then prepared. This faithfully implements the approximate transpose map achieving the highest fidelity $2/3$ on average.

{\it Minimal implementation scheme}. We now propose a general and systematic scheme of implementing the approximation $\wT$ in finite dimensions where SIC states exist. The scheme is minimal in the sense that SIC states compose spherical two-design with a minimal cardinality.

The main idea is to make use of a recent construction that a POM with $d^2$ outcomes can be decomposed into two successive POMs with $d$ outcomes each \cite{ref:kalev}. For this purpose, let us introduce two indices $k$ and $l$, each of which takes on the values $1,\ldots,d$, so that SIC states can now be denoted by $\{|s_{k,l} \rangle\}_{k,l{=}1}^{d}$. Accordingly, SIC POMs are written as, $\{{\mathfrak M}_{k,l}\}_{k,l{=}1}^{d}$ where ${\mathfrak M}_{k,l}{=} | s_{k,l} \rangle\frac1{d}\langle s_{k,l}|$. Then, once an outcome ${\mathfrak M}_{k,l}$ is obtained through the two-step measurements, the system is prepared in the corresponding state $| s_{k,l}^{*}\rangle$, according to Eq.~(\ref{eq:sict}). 

\begin{figure}[t]
\centering
\includegraphics[scale=1.]{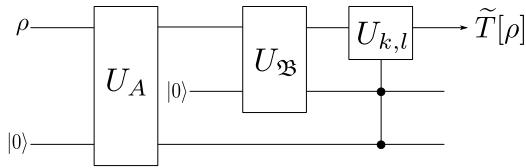}
\caption{A circuit diagram to implement the approximate transpose map is shown. Unitary transformations $U_{A}$ and $U_{\mathfrak B}$ are applied such that $\bra{k} U_{A}\ket{0}=A_k$ and $\bra{l} U_{\mathfrak B} \ket{0}={\mathfrak B}_l$ of Eq.~(\ref{eq:AB}) are implemented. Then, controlled by the forwarded information $(k,l)$, the corresponding unitary $U_{k,l}$ in Eq.~(\ref{eq:Ukl}) is applied. Finally, a quantum system initially prepared in a state $\rho$ is mapped into the state $\wT[\rho]$ at the output port.}
\label{fig:spaTrans_succ_meas}
\end{figure}

To implement the SIC POM, we first decompose it into two POMs of the form ${\mathfrak M}_{k,l}{=}A^\dagger_k {\mathfrak B}_l A_k$, where $A_k$ is the Kraus operator of the $k$th outcome of the first POM and ${\mathfrak B}_l$ is the $l$th outcome of the second POM. To this end, without loss of generality, we can safely restrict our consideration to SIC POMs that are covariant with respect to the Heisenberg-Weyl (HW) group, i.e. the so-called HW SIC POMs. This is because in all dimensions that SIC POMs are known, there exist HW SIC POMs. Then, HW SIC states can be generated by applications of the HW group elements to a fiducial state $\ket{\psi_{\textrm{fid}}}$: $| s_{k,l}\rangle = X^{k} Z^{l} |\psi_{\textrm{fid}} \rangle$ where the generalized Pauli operators $Z$ and $X$ are, $Z {=} \sum_{n=1}^{d}{\ket{n}\omega^n\bra{n}}$ and $X = \sum_{n=1}^{d}{\ket{n{\oplus} 1}\bra{n}}$ with \mbox{$\omega=e^{2\pi i/d}$} and $\oplus$ is addition modulo $d$. Assuming a fiducial vector given by $\ket{\psi_{\textrm{fid}}}{=}\sum_{n=1}^{d}\ket{n}\alpha_n$, the Kraus operators of the first POM, $A_k$, are diagonal in the computational basis where the diagonal elements given from in terms of the fiducial state probability amplitudes, while the outcomes of the second measurement, $\mathfrak{B}_{l}$, are projectors onto the Fourier-transformed basis,
\bea 
A_k = \sum_{m=1}^{d}\ket{ m\oplus k }\alpha_m\bra{m\oplus k}, \nonumber\\
\mathfrak{B}_{l} =\frac1{d} \sum_{m,n=1}^{d}\ket{ m }\omega^{(m-n)l}\bra{n}.
\label{eq:AB}
\eea
One can check that indeed the HW SIC POM outcomes are given by ${\mathfrak M}_{k,l} {=} A^\dagger_k {\mathfrak B}_l A_k$.

Once an outcome ${\mathfrak M}_{k,l}$ is obtained, the post-measurement state is given as $| s_{k,l} \rangle$, and to successfully implement $\wT$, according to Eq.~(\ref{eq:sict}), the state  $|s_{k,l}^{*}\rangle$ should then be prepared. This is accomplished by applying the unitary transformation
\bea 
U_{k,l}=X^k \Phi Z^{-2 l }X^{-k}
\label{eq:Ukl},
\eea
on the quantum system, so that $U_{k,l} |s_{k,l}\rangle=|s_{k,l}^{*}\rangle$. The diagonal operator $\Phi$ is constructed from the fiducial state: $\Phi = \mathrm{diag}[ \Phi_1,\ldots,\Phi_d]$ where $\Phi_{m} =  \alpha_{m}^{*} / \alpha_m$ for $\alpha_m{\neq}0$ and $\Phi_m =  0$ otherwise, for $m=1,\ldots,d$. In Fig.~\ref{fig:spaTrans_succ_meas}, a circuit diagram to implement the approximate transpose map is shown. The scheme provides a systematic way to optimally approximate the transpose map in a trace-preserving manner. This compares to proof-of-principle demonstrations on post-selected data reported in Ref.~\cite{ref:lim}.

\begin{figure}[t]
\centering
\includegraphics[scale=1.2]{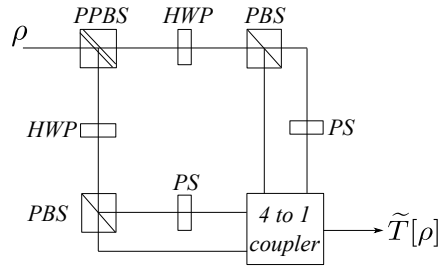}
\caption{A proposal for optical implementation of the approximate transpose is shown for polarization qubit states. See the text for details.}
\label{fig:spaTrans_qubit}
\end{figure}

As an example, we illustrate the scheme for polarization qubit states with linear optical elements. To describe the experimental setup, we refer to Fig.~\ref{fig:spaTrans_qubit}. To construct the two-step measurement of Eq.~(\ref{eq:AB}), we consider the fiducial state $|\psi_{\mathrm{fid}}\rangle = t_v |0\rangle + r_v |1\rangle$ where $t_v{=}\sqrt{3{+}\sqrt 3}/\sqrt6$, $r_v{=}e^{i\pi/4}\sqrt{3{-}\sqrt 3}/\sqrt6$, so that HW SIC POMs are generated by the action of the Pauli operators $\sigma_z$ and $\sigma_{x}$. The Kraus operators $A_{k}$ of the first measurement can be implemented by a partially-polarized beam splitter (PPBS), where the transmission ($t$) and reflection ($r$) amplitudes  for vertically ($v$) and horizontally ($h$) polarized photons are given by ($t_v$, $r_v$, $t_h$, $r_h$), respectively, with $t_h{=}r_v$, and $r_h{=}t_v$. Assuming that a photon prepared in state $\rho$ impinges the PPBS, the (unnormalized) state resulting at the transmission (reflection) port is $A_1\rho A_1^\dagger$ ($A_2\rho A_2^\dagger$). In either arm, the photon passes through a half-wave plate (HWP) placed at $22.5$ degrees to the optical axis to implement the Fourier transform on the polarization degree of freedom. The polarizing beam splitters (PBS) are then placed to transmit (reflect) horizontally (vertically) polarized photon. These implement the second measurement ${\mathfrak B}_{l}$ in the scheme of Eq.~(\ref{eq:AB}). The states of the photon at the four output ports after the PBSs are $|s_{k,l}\rangle$ with $k,l=1,2$. The probability to find the photon in one of the paths is $\tr{\rho | s_{k,l} \rangle\frac1{2}\langle s_{k,l}|}$. Then, phase shifters (PSs) that shift the polarization phase by $e^{-i\pi/4}$ are located at the appropriate path such that the preparation which transforms $|s_{k,l}\rangle$ to its complex conjugate is performed. Finally, the four output ports are combined by a 4-to-1 coupler to a single path. The state of the photon at this path is $\wT[\rho]$.

%--------------------------------------------------------------------------------------------------------------------

{\it Application to entanglement detection}. Let us now discuss how the construction and the implementation of $\wT$ can generally be applied for entanglement detection. The goal here is to translate the standard method of entanglement witnesses (EWs) into that of approximate EWs (AEWs), and to utilize the unique features of the latter. We first recall the standard scenario of EWs \cite{ref:mpr,ref:lewenstein}. A hermitian operator $W$ is an EW if $\tr{W\sigma_{\mathrm{sep}}} \geq 0$ for all separable states $\sigma_{\mathrm{sep}}$ and  $\tr{W\rho_{\mathrm{ent}}} <0$ for some entangled states $\rho_{\mathrm{ent}}$. A EW can be constructed from a positive map, $\W$, by $W = ({\mathcal I}  \otimes\W) [ Q ]$ for some operator $Q\geq0$. It suffices to consider normalized EWs, i.e. $\tr{W}=1$.  

Let $\wW$ denote the structural physical approximation of $\W$. Then, the corresponding CJ state $\rho_\wW$ is given by, $\rho_\wW = (1-p_{\rm min}) W + p_{\rm min} \openone\otimes \openone / d^2$ with minimal $p_{\rm min}$ such that $\rho_\wW\geq0$. We identify the CJ states $\rho_\wW$ as AEWs, since the relation,
\bea
\tr{\rho \rho_\wW} = (1-p_{\rm min}) \tr{\rho W} + \frac{ p_{\rm min}}{d^2}, \label{eq:newde}
\eea
shows that $\rho_\wW$ can also detect entangled states. Indeed, any entangled state $\rho_{\mathrm{ent}}$ for which $\tr{W\rho_{\mathrm{ent}}} <0$ necessarily satisfies $\text{tr}\{ \rho_\wW\rho_{\mathrm{ent}} \} {<} p_{\min}/d^2$, while for all states for which $\tr{W\rho}\geq0$, $\text{tr}\{\rho_\wW\rho\}\geq p_{\rm min}/d^2$. Therefore in terms of entanglement detection, $W$ and $\rho_\wW$ are equivalent.

\begin{figure}[t]
\centering
\includegraphics[scale=.95]{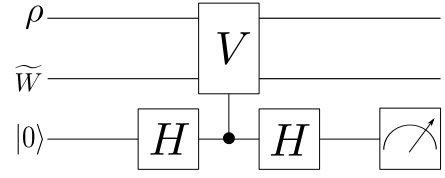}
\caption{The circuit in Ref.~\cite{ref:ekert} is exploited to estimate the quantity $\mathrm{tr}\{ \rho\rho_\wW\}$ of Eq.~(\ref{eq:newde}), with Hadamard $(H)$ and Swap $(V)$ gates. The ancillary system is measured in the computational basis and then, probabilities $p(|0\rangle) = (1 + \mathrm{tr}\{\rho\rho_\wW\} )/2$ and $p(|1\rangle) = (1 - \mathrm{tr}\{\rho\rho_\wW\} )/2$ are used to determine $\mathrm{tr}\{ \rho\rho_\wW\}$.}
\label{fig:circuit}
\end{figure}

The radical difference between EWs and AEWs lies at the fact that AEWs correspond to quantum states. First, this means that the quantity $\text{tr}\{\rho\rho_\wW\}$ of Eq.~(\ref{eq:newde}), aimed to be experimentally estimated, corresponds to the transition probability form the state $\rho$ to the CJ state  $\rho_\wW$. Therefore, experimental approaches for interference effects between quantum states would be natural implementations of estimating $\text{tr}\{\rho\rho_\wW\}$. Using the circuit that generally estimates functionals of quantum states \cite{ref:ekert}, an interferometry-based scheme is shown in Fig. \ref{fig:circuit}. Moreover, in many cases, the AEWs are separable states \cite{ref:korbicz}. To be precise, if a  positive map can detect all entangled isotropic states, then its AEW is a separable state \cite{ref:pytel, ref:remik}. Then, assuming a separable decomposition, $\rho_\wW{=} \sum_{k} q_{k} \tau_{k} \otimes \sigma_{k}$ for some $0\leq q_{k} \leq 1$, and $\sum_k q_{k}{=}1$, the detection scheme of Eq.~(\ref{eq:newde}) works with local measurement and classical communication (LOCC),
\bea
\tr{\rho\rho_\wW} = \sum_{ k} q_{k} \tr{\rho (\tau_{k} \otimes \sigma_{k})}. \label{eq:locc}
\eea
Compared to the original LOCC factorization of EWs in Ref. \cite{ref:lewen}, separable decompositions of AEWs give a natural LOCC scheme of detecting entangled states. It is, however, non-trivial in general to find the decomposition that minimize the implementation resources. This is crucial since measurement resources could also be compared to those in quantum tomography. 

Applying all these to the case of the transpose map, the parameter $p_{\rm min}$ for the approximate map is given by $p_{\rm min}{=}d/(d+1)$. As we have shown, the separable decompositions of the CJ state corresponding to $\wT$, $\rho_\wT$, follows from a spherical two-design of Eq.~(\ref{eq:spherical}). Therefore, measurements related to coherent two-designs generally provide a way to detect entangled states detected by the partial transpose map. 
%\sout{The detection scheme with corresponds  minimal resources corresponds to SIC states.}

It is straightforward to generalize AEWs to multipartite systems. Suppose that, for an $N$-partite system $C_1,\ldots,C_N$ in $S(\H_d^{\otimes N})$, we are interested to detect entanglement in a bipartite splitting $C_i$ versus the rest of the system $\bar{C}_i$, denoted as $C_i{:}\bar{C}_i$. In the original scheme, a state is entangled if $({\mathcal I}_{\bar{C}_i} \otimes {\mathcal T}_{C_i}) [\rho_{C_1 \cdots C_N}] < 0$. Now, applying the approximate map $\widetilde{\mathcal T}_{C_i}$, see Eq (\ref{eq:appt}), an AEW can be constructed as,
\bea
(\rho_\wT)_{C_i{:}\bar{C}_i}= 
\frac{1}{d^2} \sum_{k=1}^{d^2} | s_k\rangle_{C_i} \langle s_k | \otimes | \psi_k \rangle_{\bar{C}_i} \langle \psi_k |,
\eea
where $\{|s_k\rangle \}_{k=1}^{d^2}$ are SIC states in $\H_d$ and $| \psi_k \rangle{\propto}\sum_j|j\rangle_{C_1} \cdots | j \rangle_{C_{i-1}}\langle s_k|j\rangle_{C_i}|j\rangle_{C_{i+1}}\cdots|j\rangle_{C_N}$. Then we conclude that $\rho_{C_1,\ldots, C_N}$ must be entangled across a bipartition $C_i{:}\bar{C}_i$ if $\text{tr}\{\rho_{C_1,\ldots, C_N} (\rho_\wT)_{C_i{:}\bar{C}_i} \}{<}\frac{1}{d(d+1)}$.

As an example we construct an AEW to detect entanglement across a bipartite splitting of a tripartite qubit system. From the above procedure, we have that for the splitting $A{:} B C$
\bea
(\rho_\wT)_{A: B C}= 
\frac{1}{4} \sum_{k=1}^{4} | s_k\rangle_{A} \langle s_k | \otimes | \psi_k \rangle_{B C} \langle \psi_k |,\label{eq:wabc}
\eea
where $| \psi_k \rangle{\propto}\langle s_k|0\rangle|00\rangle{+}\langle s_k|1\rangle|11\rangle$. Then, a tripartite state $\rho$ is entangled in the splitting $A{:}B C$ if $\text{tr}\{\rho\rho_\wT\}{<}1/6$. Similar arguments lead to an entanglement criterion across any bipartite splitting $C_i{:}\bar{C}_i$ with $C_i{=}\{A,B,C\}$.  Consider, in particular, the tripartite state
\bea
\rho_{A,B,C} = \frac{1}{3}|\Psi  \rangle \langle \Psi  | + \frac{1}{6} ( P_{001} + P_{010} + P_{101} + P_{110} ),\nonumber
\eea
where $|\Psi\rangle$ denotes the tripartite GHZ state, $|\Psi  \rangle{\propto}|000\rangle{+}|111\rangle$, and $P_{ijk}$ is a projector onto the tripartite state $\ket{ijk}$. The state is of particular interest as it contains bound entanglement \cite{ref:dur} and also bound information \cite{ref:acin,ref:baecubitt}. Applying the AEW, one can find, 
\bea
&&\tr{\rho_{A,B,C}(\rho_\wT)_{A{:}BC} } = 1/18 < 1/6, ~\mathrm{whereas} \nonumber\\
&&\tr{\rho_{A,B,C}(\rho_\wT)_{B{:}CA} } = \tr{\rho_{A,B,C}(\rho_\wT)_{C{:}AB} }  = 1/6. \nonumber
\eea
Thus, it is shown that entanglement across the splitting $A{:}BC$ is detected.

{\it Conclusion}. In summary, we have shown that in all finite dimensions, the optimal approximation of the transpose map is given by quantum spherical two-design. This relation implies that measurement and preparation of states in coherent spherical two-design optimally approximate the transpose map. We have proposed an experimental scheme that generally and systematically implements the approximate map using SIC states. This scheme utilizes the minimal construction of the spherical two-design. We have also presented how these results can be applied to detecting multipartite entangled states. The detection scheme is based on LOCC, and is also valid in general for multipartite and high-dimensional quantum systems. Moreover, as it is shown in Fig. \ref{fig:circuit}, the detection scheme would correspond to interferometry-based experimental approaches. This envisages experimental setups estimating interferences as natural implementations of detecting entangled states. 

Finally, we should point out that while this quantitative relation between the optimal approximation of the transpose map and quantum designs has been provided here, we do not have yet an intuitive or qualitative arguments about the reason behind it or its meaning. For instance, why does state-manipulation over the symmetric subspace yields optimal approximation of the transpose map (which is the characteristics of  anti-unitary maps)? In fact, what makes more difficult and non-trivial to find a qualitative argument to the connection in general, is that in the infinite-dimension case the transpose map is no longer approximated by spherical two-designs. Note that the transpose map itself works equally well for detecting entangled states in both finite- and infinite-dimensional quantum systems \cite{ref:peres,ref:mpr,ref:duan,ref:simon}. However, while state-manipulation in coherent states optimally approximates the transpose map for Gaussian states \cite{ref:remik}, coherent states do not form two-designs and moreover Gaussian two-design does not exist \cite{ref:infinite}.

We thank R. Augusiak, A. Bendersky, D. Gross, M. Grassl, and J. Jin for valuable discussions. This research is supported by National Research Foundation and Ministry of Education (Singapore) and NSF Grants No. PHY-1212445.

%%%%%%%%%%%%%%%%%%%%%%%%%%%%%%%%%%%%%%%%%%%%%%


\begin{thebibliography}{99}


\bibitem{ref:wigner} E. Wigner, J. Math. Phys. {\bf 1}, 409 (1960).

\bibitem{ref:werner} V. Buzek,, M. Hillery, and R. F. Werner, Phys. Rev. A (R) \textbf{60} 2626 (1999).

\bibitem{ref:gisin} H. Bechmann-Pasquinucci and N. Gisin, Phys. Rev. A {\bf 59}, 4238 (1999).

\bibitem{ref:nppt} W. D\"ur, J. I. Cirac, M. Lewenstein, and D. Bruss, Phys. Rev. A \textbf{61}, 062313 (2000).

\bibitem{ref:peresconjecture} A. Peres, Found Phys. \textbf{29}, 589 (1999).

\bibitem{ref:peres} A. Peres, Phys. Rev. Lett. {\bf77}, 1413 (1996)

\bibitem{ref:mpr} M., P., and R, Horodecki, Phys. Lett. A \textbf{223} 1 (1996).

\bibitem{ref:ent-rev1} R., P., M., and K. Horodecki, Rev. Mod. Phys. \textbf{81} 865 (2009).

\bibitem{ref:ek} P. Horodecki and A. Ekert, Phys. Rev. Lett. \textbf{89}, 127902 (2002).

\bibitem{ref:spa} P. Horodecki, Phys. Rev. A \textbf{ 68} 052101 (2003).

\bibitem{ref:bus} F. Buscemi, G. M. D'Ariano, P. Perinotti, and M. F. Sacchi, Phys. Lett. A \textbf{314}, 374 (2003).

\bibitem{ref:korbicz} J. Korbicz {\it et. al.}, Phys. Rev. A \textbf{78}, 062105 (2008).

\bibitem{ref:lim} H.-T. Lim {\it et. al.}, Phys. Rev. A {\bf 83} 020301(R) (2011); H.-T. Lim {\it et. al.}, Phys. Rev. Lett. {\bf 107} 160401 (2011).

\bibitem{ref:ivanovic} I. D. Ivanovic, J. Phys. A, {\bf 14}, 3241 (1981).

\bibitem{ref:wootters89} W. K. Wootters and  B. D. Fields, Ann. Phys. (N.Y.) {\bf 191}, 363 (1989).

\bibitem{ref:sic} J. M. Renes, R. Blume-Kohout, A. J. Scott, C. M. Caves, J. Math. Phys. \textbf{45}, 2171 (2004).

\bibitem{ref:des1} G. Zauner, Int. J. Quant. Inf. {\bf 9}, 445 (2011).

\bibitem{ref:iso} A. Jamiolkowski, Rep. Math. Phys. {\bf 3}, 275 (1972); M. Choi, Linear Algebra and Its Applications, 285, (1975).

\bibitem{ref:des2} C. Dankert (MSc thesis, University of Waterloo, 2005), quant-ph/0512217; C. Dankert, R. Cleve, J. Emerson, and E. Livine, Phys. Rev. A. {\bf 80}, 012304 (2009), quant-ph/0606161. 

\bibitem{ref:des3} D. Gross, K. Audenaert, and J. Eisert, J. Math. Phys. {\bf 48}, 052104 (2007).

\bibitem{ref:des4} A. Klappenecker and M. Roetteler, quant-ph/0502031.

\bibitem{ref:scott06} A. J. Scott, J. Phys. A {\bf 39}, 13507 (2006).

\bibitem{ref:mubsic} Note that existence of MUBs, or SIC POMs, in certain dimensions has been a long-standing open problems in quantum information theory. Our results concerning about these bases are subject to the dimensions where they have been found. See also, http://qig.itp.uni-hannover.de/qiproblems/13; http://qig.itp.uni-hannover.de/qiproblems/23. 

%\bibitem{ref:grassl} A. J. Scott and M. Grassl, J. Math. Phys. \textbf{51}, 042203 (2010).

\bibitem{ref:kalev} A. Kalev, J Shang, and B. -G.  Englert, Phys. Rev. A. \textbf{85}, 052115 (2012). 

\bibitem{ref:lewenstein} M. Lewenstein, B. Kraus, J. I. Cirac, and P. Horodecki, Phys. Rev. A \textbf{62}, 052310 (2000). 

\bibitem{ref:ekert} A. Ekert {\it et. al.}, Phys. Rev. Lett. \textbf{88}, 217901 (2002).

\bibitem{ref:pytel} D. Chruscinski, J. Putel, and G. Sarbicki, Phys. Rev. A  {\bf 80}, 062314 (2009).

\bibitem{ref:remik} R. Augusiak, J. Bae, L. Czekaj, and M. Lewenstein, J. Phys. A: Math. Theor. {\bf 44}, 185308 (2011).

\bibitem{ref:lewen} O. G{\"u}hne, P. Hyllus, D. Bruss, A. Ekert, M. Lewenstein, C. Macchiavello, and A. Sanpera, J. Mod. Opt. {\bf 50}, 1079 (2003).

\bibitem{ref:dur} T. S. Cubitt, F. Verstraete, W. D\"ur, and J. I. Cirac, Phys. Rev. Lett. \textbf{91}, 037902 (2003).

\bibitem{ref:acin} A. Ac\'in, J. I. Cirac, and Ll. Masanes, Phys. Rev. Lett. \textbf{92}, 107903 (2004).

\bibitem{ref:baecubitt} J. Bae, T. Cubitt, and A. Aci\' n, Phys. Rev. A \textbf{79}, 032304 (2009).

\bibitem{ref:duan} L.-M. Duan, G. Giedke, J. I. Cirac, and P. Zoller, Phys. Rev. Lett. {\bf84}, 2722 (2000).

\bibitem{ref:simon} R. Simon, Phys. Rev. Lett. {\bf 84}, 2726 (2000).

\bibitem{ref:infinite} R. Blume-Kohout and P. S. Turner, arXiv:1110.1042.


\end{thebibliography}
\end{document}